# Unified Field-integral Thermodynamics of Bose Mixtures: Stability and Critical Behavior

Yuan-Hong Chen and Renyuan Liao[*]
*College of Physics and Energy, Fujian Normal University,*
*Fujian Provincial Key Laboratory of Quantum Manipulation and New Energy Materials, Fuzhou 350117, China and*
*Fujian Provincial Engineering Technology Research Center of Solar Energy Conversion and Energy Storage, Fuzhou 350117, China*
(Dated: August 15, 2025)

We establish a unified thermodynamic framework for Bose mixtures at finite temperatures based on the functional field integral, within which the decision on whether to discard the anomalous densities, when determining the density configuration and stability matrix, yields distinct theories. Beyond the existing Hartree-Fock approximation and Ota-Giorgini-Stringari theory, retaining the anomalous densities throughout will provide a completely self-consistent thermodynamic description, requiring the combination of the Hartree-Fock-Bogoliubov approximation and the representative statistical ensemble. Comparing three approaches for predicting magnetic susceptibility, we highlight the role of anomalous densities in stabilizing superfluid mixtures. We further unveil that Feshbach coupling can either expand the regime of atomic and molecular superfluids, or induce a phase transition to a pure molecular superfluid, depending on their density ratio. Importantly, we show that thermal fluctuations will trigger a phase transition from stable to unstable mixtures, where anomalous densities can serve as distinct signatures for experimental observation.

*Introduction*—Beyond the single-component Bose-Einstein condensates (BECs) [1, 2], bose mixtures, consisting of more than one bosonic species or hyperfine states, attract considerable interest due to their significantly richer phenomenology and more potential applications, such as phase separation (PS) [3–6], polaron excitation [7–9], quantum droplets [10–13], supersolidity [14–17] and quantized vortices [18–21]. Understanding the finite-temperature thermodynamics of bose mixtures is pivotal for unraveling their rich phase diagrams and stability conditions beyond the idealized zero-temperature limit. At zero-temperature, we focus on the quantum phase of matter and its quantum phase transition [22–24], while at finite temperatures, we are concerned with the competition between quantum fluctuations and thermal fluctuations, as well as the resulting thermodynamic behavior, such as compressibility [25–27], entropy [28–30] and specific heat [31–33]. The thermodynamics of such mixtures governs their equilibrium properties, phase diagrams, and response to external perturbations, forming a cornerstone for understanding their macroscopic behavior. While significant progress [34–40] has been made in exploring the ground-state properties of bose mixtures, a comprehensive and self-consistent thermodynamic description of interacting bose mixtures across a broad range of finite temperatures remains an outstanding challenge, particularly concerning the treatment of anomalous densities [41–43], the role of thermal fluctuations in PS, calculation of magnetic susceptibility [44–46], yet essential for connecting theory with realistic experimental conditions. Established approaches [46–54] often rely on typical approximations underestimating the effect of anomalous densities or fail to provide quantitatively accurate predictions for key thermodynamic observables.

In this Letter, we aim to bridge this gap by developing a functional field integral approach and to investigate the stability at finite temperatures and the critical behavior near phase transitions of bose mixtures. Within our framework, various theories arising from different strategies for handling anomalous densities, such as Hartree-Fock-Bogoliubov (HFB) [42, 55, 56], Ota-Giorgini-Stringari (OGS) [46, 47, 57], Hartree-Fock (HF) [1, 48, 50], and HF-like, are unified, as shown in Table I. Applying our theory to a symmetric binary Bose mixtures and comparing three approaches in Fig. 1, we reveal that for larger intraspecific interaction, the predicted instability temperature follow HF-like < HFB < OGS, while for smaller one, the order is HFB < HF-like < OGS. For the coherently coupled Bose atom-molecule mixtures [58], we find that the Feshbach coupling can extend the region of atomic and molecular superfluids (AMSF) or induce a phase transition from AMSF to molecular superfluid (MSF), as illustrated in Fig. 2. Moreover, as the temperature rises, as shown in Fig. 3, the system will undergo a magnetic phase transition [46] from stable to unstable mixtures. This phenomenon becomes more pronounced as the Feshbach coupling strength increases, and the pairing density and momentum transferring density can serve as probes.

*Framework*—Starting from the grand-canonical Hamiltonian of interacting Bose gas, we can derive the field integral representation of the grand partition function [59] $Z = \int D[\psi^*, \psi] e^{-S[\psi^*, \psi]}$ with the action $S = \int_0^\beta d\tau (\psi^* \partial_\tau \psi + H(\psi^*, \psi))$, where $\beta = 1/T$ is the inverse temperature and $\tau$ is the imaginary time. Here, we have set $\hbar = k_B = 1$ throughout and now the thermodynamic potential is the grand potential $\Omega = -T \ln Z$. When Bose-Einstein condensation (BEC) occurs, we can always decompose the field $\psi$ into a condensed part $\psi_0$ and a fluctuating part $\psi_\mathbf{k}$, with the former contributing the mean-field potential $\Omega_M$. However, the interactions between particles, both intracomponent and intercomponent, give non-Gaussian action typically, so we must adopt approximations to transform higher-order terms into quadratic one.

Although there are various seemingly chaotic procedures on the finite-temperature properties of binary Bose mixtures, they can be unified within the framework of functional integrals. As shown in Table I, such differences lie in the treatment of anomalous densities, which are directly ignored in the most well-known Popov approximation [63, 64]. Unlike single-component BECs with only one anomalous density [43], there are four ones in two-component BECs: $\sigma_{ij} = \sum_{\mathbf{k}} \langle \psi_{i,\mathbf{k}} \psi_{j,-\mathbf{k}} \rangle / V$ representing intraspecific ($i = j$) or interspecific ($i \neq j$) pairing, and $t_{12} = \sum_{\mathbf{k}} \langle \psi_{1,\mathbf{k}}^\dagger \psi_{2,\mathbf{k}} \rangle / V$ representing interspecific momentum transfer. They, in principle, together with two condensate densities $n_{i,0} = \langle \psi_{i,0}^* \psi_{i,0} \rangle / V$ and two excitation densities $n_{i,e} = \sum_{\mathbf{k}} \langle \psi_{i,\mathbf{k}}^\dagger \psi_{i,\mathbf{k}} \rangle / V$, constitute the eight order-parameters of binary BECs. Ota et al. developed a beyond mean-field description for three-dimensional homogeneous Bose mixtures at finite temperatures [46, 47, 57], which we refer to as the OGS theory. However, the Popov spirit [63] they followed results in an unphysical jump of condensate density at the BEC critical temperature, and when determining isothermal compressibility and magnetic susceptibility [47], OGS theory still artificially (or equivalently) introduces anomalous densities, leaving behind a regret of incomplete self-consistency. Thanks to the HFB approximation [42, 55] retaining all decoupling channels, it can resolve these two issues simultaneously, exhibiting the continuity at the critical temperature consistent with a second-order phase transition. Nevertheless, it should be used in conjunction with the representative statistical ensembles [60, 65, 69] proposed by Yukalov to ensure the gapless Nambu-Goldstone excitation(s), which is mathematically an application of the Hugenholtz-Pines (HP) relation(s) [70–72] and essentially a requirement of spontaneously broken U(1) symmetry [73, 74]. During such process, two chemical potentials $\mu_i^{(0)}$, $\mu_i^{(e)}$ will be introduced for each component [60–62], where one is responsible for the condensate and enters the mean-field action, and the other is responsible for the excitation and enters the Gaussian action. The former is determined through extremum conditions $\partial\Omega/\partial n_{i,0} = 0$, while the latter is derived using the HP theorem [61, 71]. The remaining extremum functions $\partial\Omega/\partial X = 0$, $X \in \{n_{i,e}, \sigma_{ij}, t_{12}\}$ form a set of self-consistent equations. After obtaining the chemical potentials, we can judge the stability by compressibility matrix [66–68]. For gases with repulsive intraspecific interactions, a positive-definite matrix allows for stable and miscible superfluids. Otherwise, in order to minimize its total energy, the system will undergo a first-order phase transition and redistribute to form an immiscible configuration or PS, where the two components occupy different domains in the available volume [75, 76] and can be determined by Maxwell construction [77] in principle. In this Letter, we continue to retain the anomalous densities when calculating the compressibility. While this greatly increases computational complexity, it is desirable and valuable to reveal the fundamental differences brought by various approaches.

TABLE I. Thermodynamic framework and method overview for Bose mixtures. OGS, Ota-Giorgini-Stringari; HFB, Hartree-Fock-Bogoliubov; HF, Hartree-Fock; HF-like, Hartree-Fock-like.

| Thermodynamic potential | | Grand potential $\Omega = -T \ln Z = \Omega_M + \Omega_G + \Omega_R$ | | |
|---|---|---|---|---|
| Density configuration determined by extremal conditions $\partial\Omega/\partial X = 0$ | How to deal with anomalous densities? | Neglect | | Take into account them and introduce two chemical potentials for each component [60–62] |
| | Overview | Popov approximation [63, 64], OGS approach [46, 47, 57] | | HFB approximation [42, 55] combined with representative ensembles [60, 65] |
| Stability determined by Hessian matrix [66–68] | How to deal with anomalous densities? | Still ignore | Introduce artificially | Continue to Retain |
| | Overview | HF-like result (HF theory [1, 48, 50]) | OGS theory [46, 47, 57] | Our approach in this Letter |

It is worth noting that in the grand potential, the contribution from Gaussian fluctuations, $\Omega_G$, can always be divided into quantum fluctuations part and thermal fluctuations one [78, 79]. For the former, the summation of energy without quasi-particle excitation, namely vacuum potential, exhibits ultraviolet divergence, so we must eliminate this irrationality [55, 80] and provide a regularization term $\Omega_R$. As one will see, the integrand in pairing densities $\sigma_{ij}$ also exhibit divergent behavior at high momentum, and the related derivative of $\Omega_R$ can accurately compensate for them [79], forming a completely self-consistent framework. The anomalous density $t_{12}$ related to momentum transfer does not suffer from the above-mentioned disasters, as it cannot spontaneously occur in a vacuum.

*Symmetric Bose mixtures*—Consider a symmetric binary Bose mixture, where two components have the same particle mass $m_i$, total density $n_i = n_{i,0} + n_{i,e}$, and intraspecific interaction strength $g_{ii} = g$, and therefore, they also have the consistent BEC critical temperature $T_{i,0} = 2\pi^2 [n_i/\zeta(3/2)]^{2/3}/m_i$ for ideal Bose gas, with $\zeta(s)$ being the Riemann zeta function. As expected, for any given interspecific interaction $g_{12}$ and temperature $T$, two gases have the equivalent non-condensed density $n_{i,e}$ and intraspecific pairing density $\sigma_{ii}$. The latter and the other two anomalous densities ($\sigma_{12}$, $t_{12}$) are ignored at first within Popov theory [47, 63] and later equivalently enter the chemical potentials [47], which is similar to assuming the complete condensation at zero-temperature and obtaining the quantum depletion [1]. Now the symmetric configuration reduces the Hessian matrix [66] with four elements to the isothermal compressibility in the total density channel and the susceptibility $\chi_T$ in the magnetization one [76], with a value of $\chi_0 = 2/(g - g_{12})$ for $T = 0$.

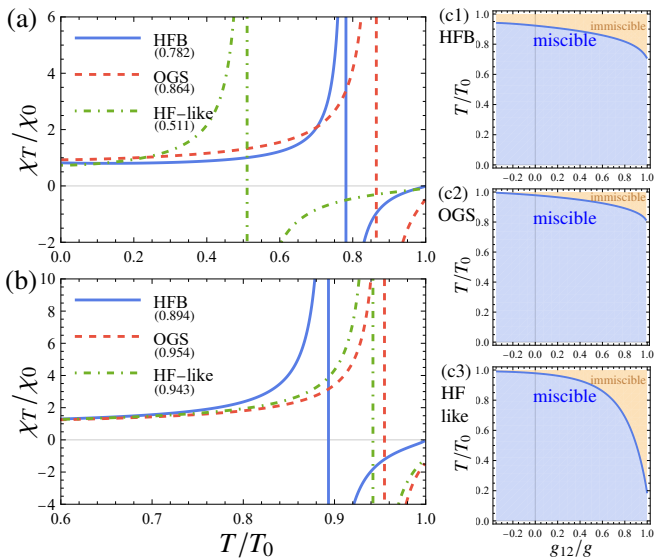

FIG. 1. Symmetrical binary Bose mixtures for the intraspecific interaction $gn = T_0/10$ with the density $n_i = n$ for each component. The magnetic susceptibility $\chi_T$ as a function of temperature $T$ for the interspecific interaction (a) $g_{12} = 9g/10$ and (b) $g_{12} = g/3$. The numbers in parentheses represent the temperature ($T_d$ in the main text) corresponding to a divergent $\chi_T$. Beyond this temperature, the susceptibility $\chi_T$ becomes negative, indicating instability. Panel (c) shows the miscibility phase diagrams, constructed by three approaches, in terms of temperature $T$ and interspecific interaction $g_{12}$. Here we omit the index $i$ of the components due to their symmetry.

Figure 1 shows the susceptibility and stability phase diagrams, where all three theories predict that the susceptibility diverges at a certain temperature, indicating dynamical instability. Above this temperature, we have $\chi_T < 0$. During the self-consistent equations solving to compressibility determination, if any anomalous density are discarded throughout, we obtain a susceptibility that is qualitatively similar to the HF approximation [1, 48, 50] and is referred to as the HF-like result. The reason for this name is that the HF approximation is actually more excessive than our treatment here, as it ignores all off-diagonal elements in the inverse Green function matrix [79], namely those terms where annihilation and creation operators appear in pairs. For larger positive interspecific interaction $g_{12}$, as shown in Fig. 1(a), the temperatures corresponding to the divergent magnetic susceptibility given by the three theories (denoted as $T_d$) are HF-like < HFB < OGS, implying the indispensable role of anomalous densities, especially interspecific densities, in mixing superfluids. The fact is that the pairing density $\sigma_{12}$ (or transferring density $t_{12}$) always enters the field integral along with $\sqrt{n_{1,0}n_{2,0}}$ term and is generally negative at higher temperature, so compared to the HFB approximation, OGS's method will provide larger absolute values of anomalous densities and larger condensate densities. Completely ignoring them like the HF approximation is not advisable, but overestimating them will result in higher $T_d$.

Nevertheless, the situation becomes subtle at smaller repulsive $g_{12}$ [Fig. 1(b)], since interspecific densities no longer dominate. They are relatively small, so there is not much difference between the critical temperatures predicted by HF-like and OGS theories for lower temperatures. Our approach (marked as HFB) will achieve a balance between intraspecific pairing densities $\sigma_{ii}$ and interspecific densities, so that the resulting $T_d$ is slightly farther from the other two. Overall, all three methods can achieve a miscible phase at lower temperature and an immiscible region at higher temperature, even for a slightly negative interspecific interaction $g_{12}$, as shown in Fig. 1(c) with the total immiscible area in the phase diagram being OGS < HF-like < HFB.

*Coherent coupled atomic-molecular superfluids*—Turn our attention to bosonic atomic-molecular mixtures. Thanks to the stimulating experimental breakthroughs in molecular BEC [81, 82], atomic and molecular superfluids (AMSF) is no longer just a fantasy. Next, we use $i = 1, 2$ to represent bosonic atoms and diatomic molecules with masses $m_2 = 2m_1$, respectively. Compared to the typical two-component Bose mixtures, at present the model includes the interconversion [58, 83] $H_F = -\alpha \sum_{\mathbf{k},\mathbf{q}} \left( \psi^\dagger_{2,\mathbf{k}+\mathbf{q}} \psi_{1,\mathbf{k}} \psi_{1,\mathbf{q}} + \text{h.c.} \right)/\sqrt{V}$ with $\alpha > 0$ being Feshbach coupling strength, in which the total atom density $n_t = n_1 + 2n_2$ is conserved.

In the absence of $\alpha$, the two condensates should be completely depleted at their respective BEC critical temperatures $T_{i,0}$, where the condensed density $n_{i,0}$ vanishes, as shown by the dashed lines in Figs. 2(a,b). Temporarily setting aside the annoying stability issue, we choose a suitable density ratio of atoms and molecules, such as $n_1 = n_2$, so that $T_{1,0} > T_{2,0}$. It is fascinating that, as shown in Fig. 2(a), the Feshbach coupling can broaden the range of AMSF up to the atomic BEC critical temperature $T_{1,0}$, where both condensed densities reach zero simultaneously. Moreover, for the same temperature $T$, Feshbach coupling leads to higher densities of condensates, and after crossing the molecular BEC critical temperature $T_{2,0}$, the decrease of its density slows down significantly as the temperature increases. These thermodynamic results are reminiscent of the coherent reaction dynamics brought about by Feshbach resonance [82, 84–86], where the atomic and molecular populations oscillate at a scale frequency after the magnetic field quenching, revealing Bose-enhanced quantum superchemistry. This suggests that after preparing AMSF in experiments, applying a constant Feshbach coupling can prolong its existence.

In the case of $T_{2,0} > T_{1,0}$, we also have similar findings shown in Fig. 2(b). However, as the temperature increases to a certain value $T_f$, the atomic condensates in the mixtures are completely depleted, and the system undergoes a second-order phase transition into a molecular superfluid (MSF) with $n_{1,0} = 0$, whose subsequent behavior is no different from that in the absence of Feshbach coupling [the red dashed line in Fig. 2(b)]. Someone consider that there might exist the atomic superfluid with $n_{2,0} = 0$, but according to the symmetry analysis [87], the effective field imposed by Feshbach coupling prohibits such possibility. Figure 2(c) presents the quantities at the AMSF–MSF phase boundary, where we find that the interspecific anomalous density ($\sigma_{12}$, $t_{12}$) vanish, yet the atomic pairing density $\sigma_{11}$ is non-zero, despite the absence of condensed atoms. This stems from the strong pairing of atoms to form molecules, which is also discussed at zero-temperature [58]. In addition, for larger ratio $n_2/n_1$ and larger Feshbach coupling $\alpha$, the phase boundary has a higher phase transition temperature $T_f$ and atomic pairing density $\sigma_{11}$, while the density of molecular condensation and pairing ($n_{2,0}$, $\sigma_{22}$) is the opposite. It indicates that we can extend the region of AMSF by enhancing the Feshbach coupling and increasing the proportion of total molecules, at the expense of achieving a smaller molecular condensed density at the final temperature.

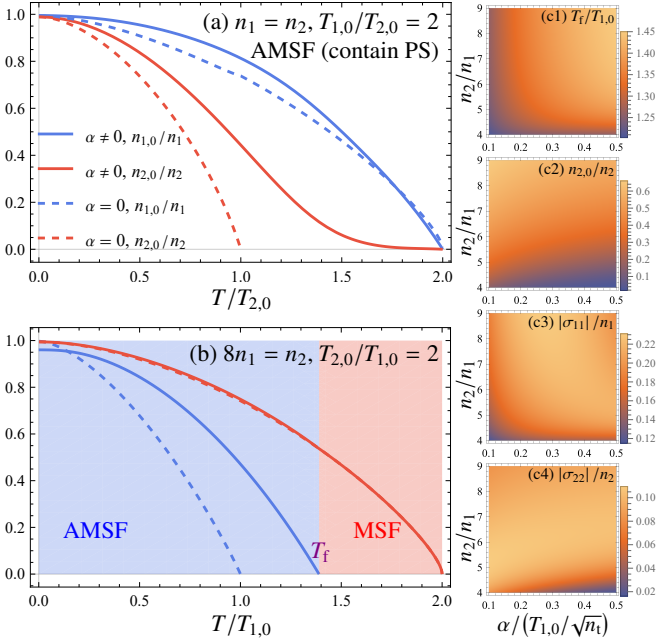

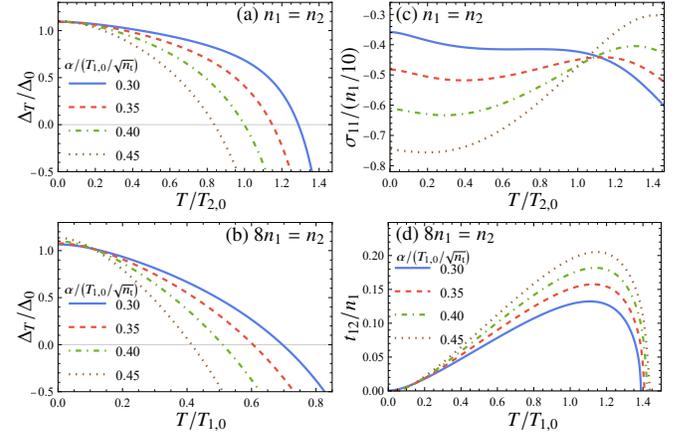

FIG. 2. Bosonic atomic-molecular mixtures with Feshbach coupling $\alpha$. (a,b) The condensate fractions $n_{i,0}/n_i$ as a function of temperature $T$, for both cases with and without $\alpha$. In (b), there is AMSF (including PS) for $0 \leq T < T_f$ and MSF for $T_f < T < 2T_{1,0}$, while in (a,b), the entire range is AMSF (including PS). (c) The quantities at the AMSF–MSF phase transition in (b), including the phase transition temperature $T_f$, molecular condensed density $n_{2,0}$, and absolute values of intraspecific pairing densities $\sigma_{ii}$, in terms of density ratio $n_2/n_1$ and Feshbach coupling strength $\alpha$. We set contact interaction parameters to $\{g_{11}, g_{22}, g_{12}\} = \{0.5, 0.4, 0.4\}$ in units of $T_{2,0}/n_t$ for (a), and in units of $T_{1,0}/n_t$ for (b,c). Take $\alpha = 0.3$ in units of $T_{2,0}/\sqrt{n_t}$ for (a), and in units of $T_{1,0}/\sqrt{n_t}$ for (b), respectively. AMSF, atomic and molecular superfluids; MSF, molecular superfluid; PS, phase separation.

FIG. 3. Stability and anomalous densities of coherent coupled atomic-molecular superfluids. (a,b) The stability criterion $\Delta_T$, in the unit of $\Delta_0$ being the mean-field result at zero-temperature, as a function of temperature $T$ for different Feshbach coupling strength $\alpha$. Positive and negative $\Delta_T$ represent stable and unstable mixture phases, respectively. (c) Atomic pairing density $\sigma_{11}$ and (d) atomic-molecular transferring density $t_{12}$ as a function of temperature $T$. The contact interaction parameters $\{g_{11}, g_{22}, g_{12}\}$ of (a,c) and (b,d) are set to the same as those of Fig. 2(a) and Fig. 2(b), respectively.

Now let's examine the outstanding stability, which is given by the compressibility matrix [66–68], equivalent to a matrix composed of the second-order derivatives of the grand potential $\Omega$. The positive definiteness of Hessian matrix serves as a stability criterion $\Delta_T$, with a constant value of $\Delta_0 = g_1 \left( g_2 + \alpha n_1 n_2^{-3/2}/2 \right) - \left( g_{12} - \alpha/\sqrt{n_2} \right)^2$ con-structed by the mean-field theory for the ground state at zero-temperature [58]. As shown in Figs. 3(a,b), the criteria $\Delta_T$ gradually decreases with increasing temperature, both for the two typical configurations mentioned earlier. The larger Feshbach coupling strength $\alpha$ provides a lower temperature corresponding to $\Delta_T = 0$. When the criteria reaches $0^+$, it is equivalent to the divergence of magnetic susceptibility [47] and can also predict the vanish of sound velocity associated with the gapless excitation spectrum [58], including a superfluid mixture of strong repulsive interspecific interaction [61] or attractive one [88]. Our parameter selection ensures that the mixture is stable at zero-temperature, so there is a possibility of magnetic phase transition with increasing temperature [46]. When the criterion $\Delta_T$ is negative, we can even explore the relationship between many-body and two-body physics, from the perspective of the transition from positive to negative in the effective scattering length [89] directly related to experiments. For temperatures $T$ below the molecular BEC critical temperature $T_{2,0}$, it is expected that a larger Feshbach coupling strength $\alpha$ gives a higher absolute atomic pairing density $\sigma_{11}$ as shown in Fig. 3(c). However, after the temperature crosses $T_{2,0}$, their relationship reverses, which can be understood as compensation for the continuation of AMSF's life, at the cost of maintaining a smaller atomic pairing. In addition, because Feshbach coupling does not directly act on molecular pairing, reflected in the fact that it does not directly enter the corresponding term in the inverse Green function matrix [79], the molecular pairing density $\sigma_{22}$ given by the different $\alpha$ are almost the same. Fig. 3(d) shows the trend of atomic-molecular momentum transferring density $t_{12}$ increasing first and then decreasing with temperature, where the increase is due to the enhancement of thermal fluctuations and the decrease is due to the response to prolong the presence of the superfluid mixture. They turn around at a temperature roughly close to $T_{1,0}$, and eventually vanish at the predicted termination temperature $T_f$ in Fig. 2(c1), indicating a transition towards MSF.

*Outlook*—Our proposed method offers a robust tool for predicting and interpreting experimental observations of Bose mixture gas at finite temperatures, and the results can be validated using path-integral Monte Carlo [90, 91]. This self-consistent description can be extended to Bose-Fermi mixtures [26] or Fermi-Fermi [92] ones by incorporating Grassmann variables [59], assessed to inhomogeneous cases through the local density approximation [93], and expanded to open quantum systems by introducing Keldysh field theory [94]. Moreover, on the basis of obtaining densities configuration, we can investigate the properties of entropy $S = -\partial \Omega / \partial T$, and further study the propagation of entropy wave, i.e., the second sound [95], in superfluid mixtures. We anticipate that our results will shed light on the interplay between quantum and thermal effects in determining stability and quantum criticality of quantum degenerate gases, revealing phenomena that are inaccessible to zero-temperature theories.

*Acknowledgments*—This work is supported by the National Natural Science Foundation of China under Grants No. 12174055 and No. 11674058, and by the Natural Science Foundation of Fujian Province under Grant No. 2020J01195.

---

* Corresponding author: ryliao@fjnu.edu.cn
[1] L. Pitaevskii and S. Stringari, *Bose-Einstein Condensation and Superfluidity* (Oxford University Press, 2016).
[2] V. I. Yukalov, Basics of Bose-Einstein condensation, Phys. Part. Nuclei **42**, 460 (2011).
[3] T.-L. Ho and V. B. Shenoy, Binary Mixtures of Bose Condensates of Alkali Atoms, Phys. Rev. Lett. **77**, 3276 (1996).
[4] T. Mishra, R. V. Pai, and B. P. Das, Phase separation in a two-species Bose mixture, Phys. Rev. A **76**, 013604 (2007).
[5] X. Jiang, S. Wu, Q. Ye, and C. Lee, Universality of miscible–immiscible phase separation dynamics in two-component Bose-Einstein condensates, New J. Phys. **21**, 023014 (2019).
[6] G. Spada, L. Parisi, G. Pascual, N. G. Parker, T. P. Billam, S. Pilati, J. Boronat, and S. Giorgini, Phase separation in binary Bose mixtures at finite temperature, SciPost Phys. **15**, 171 (2023).
[7] N. B. Jørgensen, L. Wacker, K. T. Skalmstang, M. M. Parish, J. Levinsen, R. S. Christensen, G. M. Bruun, and J. J. Arlt, Observation of Attractive and Repulsive Polarons in a Bose-Einstein Condensate, Phys. Rev. Lett. **117**, 055302 (2016).
[8] L. B. Tan, O. K. Diessel, A. Popert, R. Schmidt, A. Imamoglu, and M. Kroner, Bose Polaron Interactions in a Cavity-Coupled Monolayer Semiconductor, Phys. Rev. X **13**, 031036 (2023).
[9] N. Liu and Z. C. Tu, Polarons in binary Bose-Einstein condensates, J. Stat. Mech. **2023**, 093101 (2023).
[10] D. S. Petrov, Quantum Mechanical Stabilization of a Collapsing Bose-Bose Mixture, Phys. Rev. Lett. **115**, 155302 (2015).
[11] C. R. Cabrera, L. Tanzi, J. Sanz, B. Naylor, P. Thomas, P. Cheiney, and L. Tarruell, Quantum liquid droplets in a mixture of Bose-Einstein condensates, Science **359**, 301 (2018).
[12] T. A. Flynn, L. Parisi, T. P. Billam, and N. G. Parker, Quantum droplets in imbalanced atomic mixtures, Phys. Rev. Res. **5**, 033167 (2023).
[13] L. Cavicchioli, C. Fort, F. Ancilotto, M. Modugno, F. Minardi, and A. Burchianti, Dynamical Formation of Multiple Quantum Droplets in a Bose-Bose Mixture, Phys. Rev. Lett. **134**, 093401 (2025).
[14] J.-R. Li, J. Lee, W. Huang, S. Burchesky, B. Shteynas, F. Ç. Top, A. O. Jamison, and W. Ketterle, A stripe phase with supersolid properties in spin-orbit-coupled Bose-Einstein condensates, Nature **543**, 91 (2017).
[15] R. Liao, Searching for Supersolidity in Ultracold Atomic Bose Condensates with Rashba Spin-Orbit Coupling, Phys. Rev. Lett. **120**, 140403 (2018).
[16] R. Sachdeva, M. N. Tengstrand, and S. M. Reimann, Self-bound supersolid stripe phase in binary Bose-Einstein condensates, Phys. Rev. A **102**, 043304 (2020).
[17] W. Kirkby, A.-C. Lee, D. Baillie, T. Bland, F. Ferlaino, P. B. Blakie, and R. N. Bisset, Excitations of a Binary Dipolar Supersolid, Phys. Rev. Lett. **133**, 103401 (2024).
[18] S. T. Chui, V. N. Ryzhov, and E. E. Tareyeva, Vortex states in a binary mixture of Bose-Einstein condensates, Phys. Rev. A **63**, 023605 (2001).
[19] A. N. da Silva, R. K. Kumar, A. S. Bradley, and L. Tomio, Vortex generation in stirred binary Bose-Einstein condensates, Phys. Rev. A **107**, 033314 (2023).
[20] S. Patrick, A. Gupta, R. Gregory, and C. F. Barenghi, Stability of quantized vortices in two-component condensates, Phys. Rev. Res. **5**, 033201 (2023).
[21] A. Richaud, G. Lamporesi, M. Capone, and A. Recati, Mass-driven vortex collisions in flat superfluids, Phys. Rev. A **107**, 053317 (2023).
[22] S. L. Sondhi, S. M. Girvin, J. P. Carini, and D. Shahar, Continuous quantum phase transitions, Rev. Mod. Phys. **69**, 315 (1997).
[23] M. Heyl, Dynamical quantum phase transitions: a review, Rep. Prog. Phys. **81**, 054001 (2018).
[24] S. Sachdev, *Quantum Phases of Matter* (Cambridge University Press, Cambridge, U.K., 2023).
[25] F. J. Poveda-Cuevas, P. C. M. Castilho, E. D. Mercado-Gutierrez, A. R. Fritsch, S. R. Muniz, E. Lucioni, G. Roati, and V. S. Bagnato, Isothermal compressibility determination across Bose-Einstein condensation, Phys. Rev. A **92**, 013638 (2015).
[26] K. Manabe and Y. Ohashi, Thermodynamic stability, compressibility matrices, and effects of mediated interactions in a strongly interacting Bose-Fermi mixture, Phys. Rev. A **103**, 063317 (2021).
[27] D. Kagamihara, R. Sato, K. Manabe, H. Tajima, and Y. Ohashi, Isothermal compressibility and effects of multibody molecular interactions in a strongly interacting ultracold Fermi gas, Phys. Rev. A **106**, 033308 (2022).
[28] M. B. Kim, A. Svidzinsky, G. S. Agarwal, and M. O. Scully, Entropy of the Bose-Einstein-condensate ground state: Correlation versus ground-state entropy, Phys. Rev. A **97**, 013605 (2018).
[29] B. Yang, H. Sun, C.-J. Huang, H.-Y. Wang, Y. Deng, H.-N. Dai, Z.-S. Yuan, and J.-W. Pan, Cooling and entangling ultracold atoms in optical lattices, Science **369**, 550 (2020).
[30] P. Fabritius, J. Mohan, M. Talebi, S. Wili, W. Zwerger, M.-Z. Huang, and T. Esslinger, Irreversible entropy transport enhanced by fermionic superfluidity, Nat. Phys. **20**, 1091 (2024).
[31] R. F. Shiozaki, G. D. Telles, P. Castilho, F. J. Poveda-Cuevas, S. R. Muniz, G. Roati, V. Romero-Rochin, and V. S. Bagnato, Measuring the heat capacity in a Bose-Einstein condensation using global variables, Phys. Rev. A **90**, 043640 (2014).
[32] T. Damm, J. Schmitt, Q. Liang, D. Dung, F. Vewinger, M. Weitz, and J. Klaers, Calorimetry of a Bose-Einstein-condensed photon

gas, Nat. Commun. 7, 11340 (2016).

[33] P. van Wyk, H. Tajima, R. Hanai, and Y. Ohashi, Specific heat and effects of pairing fluctuations in the BCS-BEC-crossover regime of an ultracold Fermi gas, Phys. Rev. A 93, 013621 (2016).

[34] D. S. Hall, M. R. Matthews, J. R. Ensher, C. E. Wieman, and E. A. Cornell, Dynamics of Component Separation in a Binary Mixture of Bose-Einstein Condensates, Phys. Rev. Lett. 81, 1539 (1998).

[35] M. Pyzh and P. Schmelcher, Phase separation of a Bose-Bose mixture: Impact of the trap and particle-number imbalance, Phys. Rev. A 102, 023305 (2020).

[36] R. Liao, Ultracold Bose mixtures with spin-dependent fermion-mediated interactions, Phys. Rev. Res. 2, 043218 (2020).

[37] P. Naidon and D. S. Petrov, Mixed Bubbles in Bose-Bose Mixtures, Phys. Rev. Lett. 126, 115301 (2021).

[38] F. Isaule, I. Morera, A. Polls, and B. Juliá-Díaz, Functional renormalization for repulsive Bose-Bose mixtures at zero temperature, Phys. Rev. A 103, 013318 (2021).

[39] R. Cominotti, A. Berti, A. Farolfi, A. Zenesini, G. Lamporesi, I. Carusotto, A. Recati, and G. Ferrari, Observation of Massless and Massive Collective Excitations with Faraday Patterns in a Two-Component Superfluid, Phys. Rev. Lett. 128, 210401 (2022).

[40] L. He, H. Li, W. Yi, and Z.-Q. Yu, Quantum Criticality of Liquid-Gas Transition in a Binary Bose Mixture, Phys. Rev. Lett. 130, 193001 (2023).

[41] A. Boudjemâa and M. Benarous, Anomalous density for Bose gases at finite temperature, Phys. Rev. A 84, 043633 (2011).

[42] A. Boudjemâa, Quantum and thermal fluctuations in two-component Bose gases, Phys. Rev. A 97, 033627 (2018).

[43] A. Rakhimov and M. Nishonov, On the anomalous density of a dilute homogeneous Bose gas, Phys. Lett. A 531, 130164 (2025).

[44] T. Bienaimé, E. Fava, G. Colzi, C. Mordini, S. Serafini, C. Qu, S. Stringari, G. Lamporesi, and G. Ferrari, Spin-dipole oscillation and polarizability of a binary Bose-Einstein condensate near the miscible-immiscible phase transition, Phys. Rev. A 94, 063652 (2016).

[45] E. Fava, T. Bienaimé, C. Mordini, G. Colzi, C. Qu, S. Stringari, G. Lamporesi, and G. Ferrari, Observation of Spin Superfluidity in a Bose Gas Mixture, Phys. Rev. Lett. 120, 170401 (2018).

[46] M. Ota, S. Giorgini, and S. Stringari, Magnetic Phase Transition in a Mixture of Two Interacting Superfluid Bose Gases at Finite Temperature, Phys. Rev. Lett. 123, 075301 (2019).

[47] M. Ota and S. Giorgini, Thermodynamics of dilute Bose gases: Beyond mean-field theory for binary mixtures of Bose-Einstein condensates, Phys. Rev. A 102, 063303 (2020).

[48] P. Öhberg and S. Stenholm, Hartree-Fock treatment of the two-component Bose-Einstein condensate, Phys. Rev. A 57, 1272 (1998).

[49] B. Capogrosso-Sansone, S. Giorgini, S. Pilati, L. Pollet, N. Prokof'ev, B. Svistunov, and M. Troyer, The Beliaev technique for a weakly interacting Bose gas, New J. Phys. 12, 043010 (2010).

[50] B. V. Schaeybroeck, Weakly interacting Bose mixtures at finite temperature, Physica A 392, 3806 (2013).

[51] K. L. Lee, N. B. Jørgensen, I.-K. Liu, L. Wacker, J. J. Arlt, and N. P. Proukakis, Phase separation and dynamics of two-component Bose-Einstein condensates, Phys. Rev. A 94, 013602 (2016).

[52] L. He, P. Gao, and Z.-Q. Yu, Normal-Superfluid Phase Separation in Spin-Half Bosons at Finite Temperature, Phys. Rev. Lett. 125, 055301 (2020).

[53] A. Roy, M. Ota, A. Recati, and F. Dalfovo, Finite-temperature spin dynamics of a two-dimensional Bose-Bose atomic mixture, Phys. Rev. Res. 3, 013161 (2021).

[54] A. Roy, M. Ota, F. Dalfovo, and A. Recati, Finite-temperature ferromagnetic transition in coherently coupled Bose gases, Phys. Rev. A 107, 043301 (2023).

[55] J. O. Andersen, Theory of the weakly interacting Bose gas, Rev. Mod. Phys. 76, 599 (2004).

[56] A. Griffin, T. Nikuni, and E. Zaremba, The Beliaev and the time-dependent HFB approximations, in *Bose-Condensed Gases at Finite Temperatures* (Cambridge University Press, Cambridge, U.K., 2009) Chap. 5, pp. 81–106.

[57] M. Ota, *Sound propagation in dilute Bose gases*, phdthesis, Università di Trento, Trento, Italy (2020).

[58] Y.-H. Chen, D.-C. Zheng, and R. Liao, Quantum tricriticality of bosonic atomic-molecular mixtures with Feshbach coupling, Phys. Rev. A 110, 043321 (2024).

[59] A. Altland and B. Simons, Path Integral, in *Condensed Matter Field Theory* (Cambridge University Press, Cambridge, U.K., 2023) Chap. 3, pp. 91–170.

[60] V. I. Yukalov, Representative statistical ensembles for Bose systems with broken gauge symmetry, Ann. Phys. 323, 461 (2008).

[61] A. Rakhimov, T. Abdurakhmonov, Z. Narzikulov, and V. I. Yukalov, Self-consistent theory of a homogeneous binary Bose mixture with strong repulsive interspecies interaction, Phys. Rev. A 106, 033301 (2022).

[62] A. Sharma, G. Kartvelishvili, and J. Khoury, Finite temperature description of an interacting Bose gas, Phys. Rev. D 106, 045025 (2022).

[63] V. N. Popov, *Functional Integrals in Quantum Field Theory and Statistical Physics* (Springer Dordrecht, 1983).

[64] J. Armaitis, H. T. C. Stoof, and R. A. Duine, Hydrodynamic modes of partially condensed Bose mixtures, Phys. Rev. A 91, 043641 (2015).

[65] V. I. Yukalov and H. Kleinert, Gapless Hartree-Fock-Bogoliubov approximation for Bose gases, Phys. Rev. A 73, 063612 (2006).

[66] L. D. Landau and E. M. Lifshitz, *Statistical Physics*, 3rd ed., Course of Theoretical Physics, Vol. 5 (Pergamon, 1980).

[67] L. Viverit, C. J. Pethick, and H. Smith, Zero-temperature phase diagram of binary boson-fermion mixtures, Phys. Rev. A 61, 053605 (2000).

[68] H. Shi, W.-M. Zheng, and S.-T. Chui, Phase separation of Bose gases at finite temperature, Phys. Rev. A 61, 063613 (2000).

[69] V. I. Yukalov and E. P. Yukalova, Bose-Einstein-condensed gases with arbitrary strong interactions, Phys. Rev. A 74, 063623 (2006).

[70] N. M. Hugenholtz and D. Pines, Ground-State Energy and Excitation Spectrum of a System of Interacting Bosons, Phys. Rev. 116, 489 (1959).

[71] S. Watabe, Hugenholtz-Pines theorem for multicomponent Bose-Einstein condensates, Phys. Rev. A 103, 053307 (2021).

[72] A. Rakhimov and A. Khudoyberdiev, Hugenholtz-Pines relations and the critical temperature of a Rabi coupled binary Bose system, Eur. Phys. J. D 77, 37 (2023).

[73] J. Goldstone, A. Salam, and S. Weinberg, Broken Symmetries, Phys. Rev. 127, 965 (1962).

[74] H. Enomoto, M. Okumura, and Y. Yamanaka, Goldstone theorem, Hugenholtz-Pines theorem, and Ward-Takahashi relation in finite volume Bose-Einstein condensed gases, Ann. Phys. 321, 1892 (2006).

[75] S. B. Papp, J. M. Pino, and C. E. Wieman, Tunable Miscibility in a Dual-Species Bose-Einstein Condensate, Phys. Rev. Lett. 101, 040402 (2008).

[76] C. Baroni, G. Lamporesi, and M. Zaccanti, Quantum mixtures of ultracold gases of neutral atoms, Nat. Rev. Phys. 6, 736 (2024).

[77] L. E. Reichl, The Thermodynamics of Phase Transitions, in *A

*Modern Course in Statistical Physics* (Wiley-VCH, Weinheim, Ger, 2016) Chap. 4, pp. 87–134.

[78] R. Liao, Z.-G. Huang, X.-M. Lin, and O. Fialko, Spin-orbit-coupled Bose gases at finite temperatures, Phys. Rev. A **89**, 063614 (2014).

[79] See Supplemental Material for the mathematical formalism of thermodynamic framework and its applications in single-component BECs, binary Bose mixtures, and coherent coupled atomic-molecular superfluids.

[80] H. T. C. Stoof, K. B. Gubbels, and D. B. M. Dickerscheid, *Ultracold Quantum Fields* (Springer Dordrecht, 2009).

[81] Z. Zhang, L. Chen, K.-X. Yao, and C. Chin, Transition from an atomic to a molecular Bose-Einstein condensate, Nature **592**, 708 (2021).

[82] Z. Zhang, S. Nagata, K.-X. Yao, and C. Chin, Many-body chemical reactions in a quantum degenerate gas, Nat. Phys. **19**, 1466 (2023).

[83] M. Bellitti, G. Goldstein, and C. R. Laumann, Lifetime of excitations in atomic and molecular Bose-Einstein condensates, Phys. Rev. A **107**, 033304 (2023).

[84] D. Pimenov and E. J. Mueller, Bose-enhanced relaxation of driven atom-molecule condensates, Phys. Rev. A **109**, 043311 (2024).

[85] Z. Wang, K. Wang, Z. Zhang, S. Nagata, C. Chin, and K. Levin, Stability and dynamics of atom-molecule superfluids near a narrow Feshbach resonance, Phys. Rev. A **110**, 013306 (2024).

[86] K. Wang, Z. Zhang, S. Nagata, Z. Wang, and K. Levin, Universal coherent atom-molecule oscillations in the dynamics of the unitary Bose gas near a narrow Feshbach resonance, Phys. Rev. Res. **7**, L012025 (2025).

[87] L. Radzihovsky, P. B. Weichman, and J. I. Park, Superfluidity and phase transitions in a resonant Bose gas, Ann. Phys. **323**, 2376 (2008).

[88] A. Rakhimov, S. Tukhtasinova, and V. I. Yukalov, Stability of binary Bose mixtures with attractive intercomponent interactions, arXiv:2505.21108 [cond-mat.quant-gas] (2025).

[89] Z. Wang, K. Wang, Z. Zhang, Q. Chen, C. Chin, and K. Levin, Tunable Molecular Interactions Near an Atomic Feshbach Resonance: Stability and Collapse of a Molecular Bose-Einstein Condensate, arXiv:2504.09183 [cond-mat.quant-gas] (2025).

[90] K. Dželalija, V. Cikojević, J. Boronat, and L. Vranješ Markić, Trapped Bose-Bose mixtures at finite temperature: A quantum Monte Carlo approach, Phys. Rev. A **102**, 063304 (2020).

[91] G. Spada, S. Pilati, and S. Giorgini, Thermodynamics of a dilute Bose gas: A path-integral Monte Carlo study, Phys. Rev. A **105**, 013325 (2022).

[92] Q. Chen, Y. He, C.-C. Chien, and K. Levin, Stability conditions and phase diagrams for two-component Fermi gases with population imbalance, Phys. Rev. A **74**, 063603 (2006).

[93] F. Dalfovo, S. Giorgini, L. P. Pitaevskii, and S. Stringari, Theory of Bose-Einstein condensation in trapped gases, Rev. Mod. Phys. **71**, 463 (1999).

[94] L. M. Sieberer, M. Buchhold, J. Marino, and S. Diehl, Universality in driven open quantum matter, Rev. Mod. Phys. **97**, 025004 (2025).

[95] H. Hu, X.-C. Yao, and X.-J. Liu, Second sound with ultracold atoms: a brief review, AAPPS Bull. **32**, 26 (2022).

# Supplemental Material for
# Unified Field-integral Thermodynamics of Bose Mixtures: Stability and Critical Behavior

Yuan-Hong Chen and Renyuan Liao [*]
*College of Physics and Energy, Fujian Normal University,
Fujian Provincial Key Laboratory of Quantum Manipulation and New Energy Materials, Fuzhou 350117, China and
Fujian Provincial Engineering Technology Research Center of Solar Energy Conversion and Energy Storage, Fuzhou 350117, China*
(Dated: August 14, 2025)

In this supplemental material, we demonstrate the mathematical formalism of the self-consistent thermodynamic framework based on functional field integral, including our approach of combining Hartree-Fock-Bogoliubov approximation with representative statistical ensemble. We first examine it on the well-known single-component interacting Bose gas, showing the complete self-consistency. We also apply it to binary Bose mixtures and atomic-molecular superfluids, supplement the former's condensed density and pairing density with temperature, and provide the specific expressions of each part of the thermodynamic potential.

*Formalism of thermodynamic framework*—Taking reasonable approximation, the total action $S$ can be divided into the mean-field contribution $S_M$ and the Gaussian fluctuations term $S_G$. Choosing the appropriate base vector $\Phi$, $S_G$ can be written as $S_G = \left(\sum_k \Phi^\dagger \mathcal{G}^{-1}\Phi - \beta\,\{\text{tfcr}\}\right)/2$, where we define the following notations: $k \equiv (\mathbf{k}, i\omega_n)$ with the momentum $\mathbf{k}$ and the Matsubara frequency $\omega_n = 2n\pi T$, $\mathcal{G}$ is the Green function, $\beta = 1/T$ is the inverse temperature, and $\{\text{tfcr}\}$ denotes those terms from commutation relation of Bose operators. We set $\hbar = k_B = 1$ throughout. Deriving from the partition function $Z = \int D[\psi^*, \psi] e^{-S[\psi^*,\psi]}$, we obtain the grand potential

$$\Omega = -T \ln Z = T S_M + \frac{\text{Tr}\left(\ln\left(\beta \mathcal{G}^{-1}\right)\right)}{2\beta} - \frac{\{\text{tfcr}\}}{2}, \quad (1)$$

of which the latter two constitute the Gaussian fluctuations potential $\Omega_G$. Solving the secular equation $\det \mathcal{G}^{-1}(i\omega_n, \mathbf{k}) = 0$, we can determine the excitation spectrum(s) $\mathcal{E}_s(\mathbf{k})$ with the label $s$, so that we have

$$\Omega_G = T \sum_{s,\mathbf{k}\neq 0} \ln\left[1 - \exp\left(-\frac{\mathcal{E}_s}{T}\right)\right] + \sum_{s,\mathbf{k}\neq 0} \frac{\mathcal{E}_s - \{\text{tfcr}\}}{2}, \quad (2)$$

where the former comes from thermal fluctuations and the latter from quantum fluctuations. As stated in the main text, it is necessary to insert the regularization term $\Omega_R$ to eliminate the ultraviolet divergence. So far, the total thermodynamic potential is $\Omega = \Omega_M + \Omega_G + \Omega_R$ with the mean-field part $\Omega_M = T S_M$.

Bose mixtures generally have these densities: condensed density $n_{i,0}$, non-condensed density $n_{i,e}$, pairing density $\sigma_{ij}$, and momentum transfering density $t_{ij}$. Solving the saddle-point equation $\partial\Omega/\partial n_{i,0} = 0$, we can determine $\mu_i^{(0)}$, the chemical potential responsible for the condensate. The remaining extremum conditions $\partial\Omega/\partial X = 0$, $X \in \{n_{i,e}, \sigma_{ij}, t_{ij}\}$ will provide a set of self-consistent equations $X = f(X)$, which determine the ground-state densities configuration together with the particle-number conserving $N_i = (n_{i,0} + n_{i,e}) V$. The stability of the system is judged by the positive definiteness of $\partial^2\Omega/\partial n_{i,0}^2$ or the matrix formed by them.

*Application I: Single-component interacting Bose gas*—The Hamiltonian is $\hat{H} = \sum_\mathbf{k} \varepsilon_\mathbf{k} \hat{\psi}_\mathbf{k}^\dagger \hat{\psi}_\mathbf{k} + \frac{g}{2V}\sum_{\mathbf{k}+\mathbf{p}=\mathbf{q}+\mathbf{m}} \hat{\psi}_\mathbf{k}^\dagger \hat{\psi}_\mathbf{p}^\dagger \hat{\psi}_\mathbf{q} \hat{\psi}_\mathbf{m}$, where we have notations: $\varepsilon_\mathbf{k} = \mathbf{k}^2/(2m)$ single-particle dispersion, $m$ particle mass, $\hat{\psi}_\mathbf{k}^\dagger$ creation operator with momentum $\mathbf{k}$, $\hat{\psi}_\mathbf{q}$ annihilation operator with momentum $\mathbf{q}$, $g$ repulsive contact interaction strength, $V$ system volume.

Applying Hartree-Fock-Bogoliubov (HFB) approximation

$$\begin{aligned}
&\psi_\mathbf{k}^\dagger \psi_\mathbf{p}^\dagger \psi_\mathbf{q} \psi_\mathbf{m} \delta_{\mathbf{k}+\mathbf{p},\mathbf{q}+\mathbf{m}} \\
&\simeq 4\psi_\mathbf{k}^\dagger \psi_\mathbf{k} \left\langle \psi_\mathbf{p}^\dagger \psi_\mathbf{p} \right\rangle + \left(\psi_\mathbf{k}^\dagger \psi_{-\mathbf{k}}^\dagger \left\langle \psi_\mathbf{q}\psi_{-\mathbf{q}} \right\rangle + \text{h.c.}\right) \\
&\quad - 2\left\langle \psi_\mathbf{k}^\dagger \psi_\mathbf{k}\right\rangle \left\langle \psi_\mathbf{p}^\dagger \psi_\mathbf{p} \right\rangle - \left\langle \psi_\mathbf{k}^\dagger \psi_{-\mathbf{k}}^\dagger \right\rangle \left\langle \psi_\mathbf{q}\psi_{-\mathbf{q}} \right\rangle
\end{aligned} \quad (3)$$

to the untreated Hamiltonian and introducing two chemical potentials ($\mu^{(0)}$, $\mu^{(e)}$) in the representative statistical ensemble, we obtain the quadratic grand-canonical Hamiltonian for single-component Bose superfluid $\hat{H}_\mu = \Omega_M + \sum_{\mathbf{k}\neq 0}\xi_\mathbf{k}\psi_\mathbf{k}^\dagger\psi_\mathbf{k} + \mathcal{M}_{12}\sum_{\mathbf{k}\neq 0}(\psi_\mathbf{k}\psi_{-\mathbf{k}} + \text{h.c.})/2$ with the mean-field energy $\Omega_M = \left(\frac{gn_0^2}{2} - gn_e^2 - \frac{g\sigma^2}{2} - \mu^{(0)}n_0\right)V$, so that the inverse Green function is written as $\mathcal{G}^{-1} = \begin{pmatrix} -i\omega_n + \xi_\mathbf{k} & \mathcal{M}_{12} \\ \mathcal{M}_{12} & i\omega_n + \xi_\mathbf{k} \end{pmatrix}$, with $\xi_\mathbf{k} = \varepsilon_\mathbf{k} + 2gn - \mu^{(e)}$ being the effective single-particle kinetic energy and $\mathcal{M}_{12} = g(n_0 + \sigma)$ being the pairing matrix element, and we assume that pairing density $\sigma$ is real without loss of generality. Using the Hugenholtz-Pines (HP) theorem [1], which is equivalent to requiring the excitation spectrum $\mathcal{E}(\mathbf{k})$ to be gapless, we can determine the chemical potential responsible for the depletion is $\mu^{(e)} = g(n_0 + 2n_e - \sigma)$. The zero-point energy $\sum_{\mathbf{k}\neq 0}[\mathcal{E}(\mathbf{k}) - \mathcal{M}_{12}]/2$, caused by quantum fluctuations, needs to employ renormalization, so that the Gaussian fluctuations potential develops into $\Omega_G \to \Omega_G + \Omega_R$, where $\Omega_R = \sum_{\mathbf{k}\neq 0} \frac{\mathcal{M}_{12}^2/2}{\mathbf{k}^2/(2m_r)}$ is the regularization term. Introducing the effective mass $m_r = m/2$ for two-body collision will better reflect its meaning in the case of multi-component. The saddle-point equation $\partial\Omega/\partial\sigma = 0$ provides a self-consistent equation about $\sigma$:

$$\sigma = \sum_{\substack{\mathbf{k}\neq 0 \\ i\omega_n}} \frac{\mathcal{G}_{12}}{\beta V} + \frac{1}{gV}\frac{\partial}{\partial\sigma}\sum_{\mathbf{k}\neq 0}\frac{\mathcal{M}_{12}^2/2}{\mathbf{k}^2/(2m_r)}, \quad (4)$$

where $\mathcal{G}_{12}$ is a function of $\sigma$ and the latter term exactly offsets the ultraviolet divergence of the integrand in the former term.

Similarly, we have $n_e = \sum \mathcal{G}_{11}/(\beta V)$ for the depletion density. In addition, the extremum condition $\partial\Omega/\partial n_0 = 0$ gives $\mu^{(0)} = g(n_0 + 2n_e + \sigma)$, which is closely related to the fact that $\sigma$ and $n_0$ always enter $\mathcal{M}_{12}$ together. As stated in the main text, if the anomalous density $\sigma$ is ignored, the Popov result or the Ota-Giorgini-Stringari (OGS) theory [2, 3] will be obtained. If only HFB approximation is adopted without representative statistical ensemble, the excitation spectrum still has energy gap. Finally, stability is determined by the compressibility

$$(\kappa_T)^{-1} = gV + \frac{1}{2\beta}\mathrm{Tr}\left[\mathcal{G}\frac{\partial^2\mathcal{G}^{-1}}{\partial n_0^2} - \left(\mathcal{G}\frac{\partial\mathcal{G}^{-1}}{\partial n_0}\right)^2\right] + \frac{\partial^2\Omega_R}{\partial n_0^2}. \quad (5)$$

The divergent behavior of the *Trace* in middle and the second derivative of $\Omega_R$ cancel each other out. For a single-component Bose-Einstein condensates (BECs), $\kappa_T$ is always positive, but it diverges near the critical BEC temperature $T_0 = 2\pi^2 [n/\zeta(3/2)]^{2/3}/m$ for the ideal Bose gas.

*Application II: Binary Bose mixtures*—The Hamiltonian $\hat{H} = \hat{H}_1 + \hat{H}_2 + \hat{H}_{12}$ contains the single-particle kinetic energy plus their intraspecific contact repulsive interaction $\hat{H}_i = \sum_{\mathbf{k}} \varepsilon_{i,\mathbf{k}} \hat{\psi}_{i,\mathbf{k}}^\dagger \hat{\psi}_{i,\mathbf{k}} + \frac{g_i}{2V}\sum_{\mathbf{k+p=q+m}} \hat{\psi}_{i,\mathbf{k}}^\dagger \hat{\psi}_{i,\mathbf{p}}^\dagger \hat{\psi}_{i,\mathbf{q}} \hat{\psi}_{i,\mathbf{m}}$ and the interspecific interaction $\hat{H}_{12} = g_{12} \sum_{\mathbf{k+q=p+m}} \hat{\psi}_{1,\mathbf{k}}^\dagger \hat{\psi}_{1,\mathbf{p}} \hat{\psi}_{2,\mathbf{q}}^\dagger \hat{\psi}_{2,\mathbf{m}}/V$. Making use of the Bogoliubov decomposition, applying HFB approximation, i.e. Eq. (3) and

$$\frac{1}{V}\sum_{\mathbf{p,q,m}} \psi_{1,\mathbf{k}}^\dagger \psi_{1,\mathbf{p}} \psi_{2,\mathbf{q}}^\dagger \psi_{2,\mathbf{m}} \delta_{\mathbf{k+q,p+m}}$$
$$\simeq \sum_{i=1,2} n_{i,e} \psi_{3-i,\mathbf{k}}^\dagger \psi_{3-i,\mathbf{k}} - \left(n_{1,e}n_{2,e} + \sigma_{12}^2 + t_{12}^2\right)V$$
$$+ \left[\left(t_{12}\psi_{2,\mathbf{k}}^\dagger \psi_{1,\mathbf{k}} + \sigma_{12}\psi_{1,\mathbf{k}}^\dagger \psi_{2,-\mathbf{k}}^\dagger\right) + \mathrm{h.c.}\right], \quad (6)$$

to Hamiltonian, and finally introducing two chemical potentials for each component, we obtain the mean-field potential

$$\frac{\Omega_M}{V} = \sum_{i=1,2} \left(\frac{g_i n_{i,0}^2}{2} - g_i n_{i,e}^2 - \frac{g_i \sigma_{ii}^2}{2} - \mu_i^{(0)} n_{i,0}\right)$$
$$+ g_{12}\left(n_{1,0}n_{2,0} - n_{1,e}n_{2,e} - \sigma_{12}^2 - t_{12}^2\right). \quad (7)$$

With $\Phi = \left(\psi_{1,k}, \psi_{1,-k}^*, \psi_{2,k}, \psi_{2,-k}^*\right)^T$ as the base vector, the inverse Green function in the grand-canonical ensemble can be written as

$$\mathcal{G}^{-1} = \begin{pmatrix} -i\omega_n + \xi_{1,\mathbf{k}} & \mathcal{M}_{12} & \mathcal{T} & \mathcal{P} \\ \mathcal{M}_{12} & i\omega_n + \xi_{1,\mathbf{k}} & \mathcal{P} & \mathcal{T} \\ \mathcal{T} & \mathcal{P} & -i\omega_n + \xi_{2,\mathbf{k}} & \mathcal{M}_{34} \\ \mathcal{P} & \mathcal{T} & \mathcal{M}_{34} & i\omega_n + \xi_{2,\mathbf{k}} \end{pmatrix},$$

where we have defined the following matrix elements: the effective single-particle kinetic energy $\xi_{i,\mathbf{k}} = \varepsilon_{i,\mathbf{k}} + (2g_i n_i + g_{12}n_{3-i}) - \mu_i^{(e)}$, intraspecific pairing potential $\mathcal{M}_{12} = g_1(n_{1,0} + \sigma_{11})$ and $\mathcal{M}_{34} = g_2(n_{2,0} + \sigma_{22})$, interspecific pairing $\mathcal{P} = g_{12}\left(\sqrt{n_{1,0}n_{2,0}} + \sigma_{12}\right)$, and interspecific momentum transfering $\mathcal{T} = g_{12}\left(\sqrt{n_{1,0}n_{2,0}} + t_{12}\right)$. Similar to the one-component BEC, from the extremum conditions $\partial\Omega/\partial n_{i,0} = 0$, we can obtain the chemical potentials $\mu_i^{(0)} = g_i\left(n_i + n_{i,e} + \sigma_{ii}\right) + g_{12}\left(n_{3-i} + (t_{12} + \sigma_{12})\sqrt{n_{3-i,0}/n_{i,0}}\right)$ responsible for the condensates. According to the HP theorem for multi-component BECs [4], we have $\mu_i^{(e)} = g_i\left(n_{i,0} + 2n_{i,e} - \sigma_{ii}\right) + g_{12}n_{3-i}$ and $t_{12} = \sigma_{12}$. The former ensures that when interspecific interaction is absent, the two decoupled excitation spectra are gapless, while the latter further ensures that when interspecific interaction exists, they are gapless at the same time, resulting from the spontaneous breaking of U(1) × U(1) symmetry.

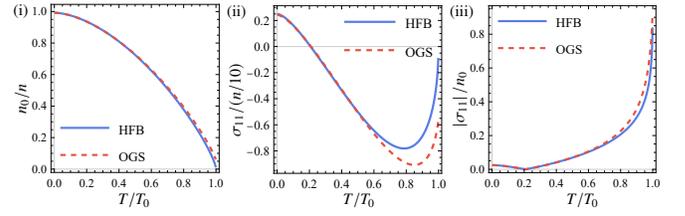

FIG. S1. Symmetric Bose binary mixtures with component index ignored. (i) Condensed fraction $n_0/n$, (ii) intraspecific pairing density $\sigma_{11}$ in units of the total density $n$, and (iii) the ratio of the absolute value of pairing density to condensed density as a function of temperature $T$. The parameters are the same as Fig. 1(a) in the main text.

The analysis of asymptotic behavior of quantum fluctuations provides the regularization potential

$$\Omega_R = \sum_{\mathbf{k}\neq 0}\frac{1}{\mathbf{k}^2}\left(\frac{m_1\mathcal{M}_{12}^2 + m_2\mathcal{M}_{34}^2}{2} + 2m_R\mathcal{P}^2\right) \quad (8)$$

with $m_R = m_1 m_2/(m_1 + m_2)$ being the effective mass in interspecific scattering, and its corresponding derivatives play an indispensable role in determining the densities configuration $\{n_{i,0}, n_{i,e}, \sigma_{ij}, t_{12}\}$ and the compressibility matrix, showing the perfect self-consistency of our framework.

In a symmetric configuration, i.e. $m_1 = m_2$ for mass, $g_1 = g_2$ for intraspecific interaction strength, $n_1 = n_2$ for the total density of gas, the system has the same BEC critical temperature $T_{i,0}$, condensed density $n_{i,0}$ and intraspecific pairing density $\sigma_{ii}$, for a given temperature $T$ and interspecific interaction strength $g_{12}$. As shown in Fig. S1(i), compared with the OGS theory, our scheme (labeled HFB) achieves a smaller condensed density $n_0$ at the same temperature, which is due to the competition between condensed density and pairing density. Furthermore, it exhibits the continuity at the BEC critical temperature $T = T_0$, consistent with a second-order phase transition obeying a spontaneous breaking of symmetry. Fig. S1(ii) shows that the intraspecific pairing density is generally negative at higher temperatures, and the OGS theory will obtain results with larger absolute values. There are similar conclusions for the interspecific pairing density. Whether it is the OGS approach or our method that always retains the pairing density in this paper, we can witness that, except in the extremely low temperature, where the quantum fluctuations suppress the thermal fluctuations, the proportion of the absolute value of pairing density to the density of condensation increases with the rise of temperature, as shown in Fig. S1(iii). It can be seen that in the region dominated by thermal fluctuations, we can no longer belittle the role of pairing densities in condensates and determining its stability.

*Application III: Coherent coupled atomic-molecular superfluids*—Now the Hamiltonian for the atomic-molecular mixtures is $\hat{H} = \sum_{i=1,2} \hat{H}_i + \hat{H}_{12} + \hat{H}_F$ with Feshbach coupling $\hat{H}_F = -\alpha \sum_{\mathbf{k},\mathbf{q}} \left( \hat{\psi}^\dagger_{2,\mathbf{k}+\mathbf{q}} \hat{\psi}_{1,\mathbf{k}} \hat{\psi}_{1,\mathbf{q}} + \text{h.c.} \right)/\sqrt{V}$, where $i = 1, 2$ represent atoms and diatomic molecules, respectively.

Because the Feshbach coupling is an interaction of three operators, it does not produce any new terms in the HFB approximation, and the change to the action comes from the standard Bogoliubov split. Therefore, we only need to replace the mean-field potential $\Omega_M$ in the binary Bose mixtures with $\Omega_M - 2\alpha n_{1,0}\sqrt{n_{2,0}}$, and make a transformation $\mathcal{M}_{12} \to \mathcal{M}_{12} - \mathcal{A}_2$ and $\mathcal{T} \to \mathcal{T} - \mathcal{A}_1$ in the inverse Green function $\mathcal{G}$, where $\mathcal{A}_i = 2\alpha\sqrt{n_{i,0}}$ are effective coupling of interspecific momentum transfer and atomic pairing, respectively. Furthermore, the Feshbach coupling locks the phase of the two condensates, so that the symmetry of the order parameters is reduced from $U(1) \times U(1)$ to $U(1) \times \mathbb{Z}_2$ [5, 6]. When the molecules are condensed yet atoms are not, the system is in the molecular superfluid (MSF) phase with spontaneous breaking of $U(1)$ symmetry. If the atoms are further condensed, the system is atomic and molecular superfluids (AMSF), and the remaining $\mathbb{Z}_2$ symmetry is broken. Therefore, in the coherent coupled superfluids, there is a gapless excitation corresponding to in-phase fluctuations of two phases and a gapped excitation corresponding to out-of-phase fluctuations. On this basis, we can determine $\mu_i^{(e)} = g_i \left( n_{i,0} + 2n_{i,e} - \sigma_{ii} \right) + g_{12} n_{3-i} - \mathcal{A}_{2i}$ with $\mathcal{A}_4 = \mathcal{A}_1^2/(2\mathcal{A}_2)$. As mentioned in the main text, the fact is that an atomic superfluid does not exist, ensuring that $n_{2,0}$ in the denominator is not troublesome.

* Corresponding author: ryliao@fjnu.edu.cn
[1] V. I. Yukalov, Representative statistical ensembles for Bose systems with broken gauge symmetry, Ann. Phys. **323**, 461 (2008).
[2] M. Ota, S. Giorgini, and S. Stringari, Magnetic Phase Transition in a Mixture of Two Interacting Superfluid Bose Gases at Finite Temperature, Phys. Rev. Lett. **123**, 075301 (2019).
[3] M. Ota and S. Giorgini, Thermodynamics of dilute Bose gases: Beyond mean-field theory for binary mixtures of Bose-Einstein condensates, Phys. Rev. A **102**, 063303 (2020).
[4] S. Watabe, Hugenholtz-Pines theorem for multicomponent Bose-Einstein condensates, Phys. Rev. A **103**, 053307 (2021).
[5] Y.-H. Chen, D.-C. Zheng, and R. Liao, Quantum tricriticality of bosonic atomic-molecular mixtures with Feshbach coupling, Phys. Rev. A **110**, 043321 (2024).
[6] L. Radzihovsky, P. B. Weichman, and J. I. Park, Superfluidity and phase transitions in a resonant Bose gas, Ann. Phys. **323**, 2376 (2008).